\documentclass[twocolumn,prb]{revtex4}
\usepackage{epsfig,amssymb,amsmath}
\begin{document}
\title{Dynamics of DC-SQUID with nontrivial barriers under external radiation}
\author{I. R. Rahmonov~$^{1,2}$}
\author{Yu. M. Shukrinov~$^{1,3}$}
\author{R. Dawood~$^{4}$}
\author{P. Seidel~$^{5}$}
\author{K. Sengupta~$^{6}$}

\address{$^{1}$ BLTP, JINR, Dubna, Moscow Region, 141980, Russia\\
$^{2}$Umarov Physical Technical Institute, TAS, Dushanbe, 734063, Tajikistan\\
$^{3}$ Dubna State University, Dubna,  141980, Russia\\
$^{4}$ Cairo University, Giza, 12613, Egypt\\
$^{5}$ Institut f\"ur Festk\"orperphysik, Friedrich Schiller Universit\"at - Jena, D-07743 Jena, Germany\\
$^{6}$ Indian Association for the Cultivation of Science, Kolkata, 700 032, India.
}

\date{\today}

\begin{abstract}
	We study the phase dynamics and IV--characteristics of DC--SQUIDs consisting of Josephson junctions with topologically nontrivial barriers, which cause the appearance of Majorana bound state. Its comparative analysis with the trivial case is performed. The influence of external electromagnetic radiation is considered and the analysis of the amplitude dependence of the Shapiro step widths is performed. We have shown that in nontrivial case the width of even harmonic of Shapiro step is larger than width of odd harmonic. In the presence of external dc magnetic field a beating state is realized in the DC--SQUID, which leads to a resonance branch in the IV--curve. We show that in the presence of resonance branch the maximum width of Shapiro step and periods of its amplitude dependence are decreased in comparison of situation without resonance in both trivial and nontrivial cases. We demonstrate that in the presence of the resonance branch the chaotic behavior of IV--curve is reduced.
\vspace{2cm}


\end{abstract}
\maketitle
\section*{INTRODUCTION}
Recently, there have been many attempts to develop a quantum computer based on the Majorana fermions~\cite{majorana37}, which are predicted to exist in Josephson junctions (JJ) with topologically nontrivial barriers (TNB)~\cite{fu08,tanaka09}. It is assumed that nontrivial state is formed at the boundary or on the surface of the topological insulator~\cite{fu07} and a semiconductor nanowire in the presence of Rashba spin -- orbit coupling and Zeeman field~\cite{sau10}. SQUIDs with TNB can be used for detecting and controlling the Majorana fermions~\cite{fu09}. The formation of Majorana states in JJ leads to the tunneling of quasiparticles with charge $e$, which doubles the Josephson current period~\cite{fu09,kitaev01} $I_{s}=I_{c}\sin{\varphi/2}$.

In the presence of external magnetic field a branch appears in the IV characteristics corresponding to the resonance frequency of the Josephson and electromagnetic oscillations~\cite{schmidt85}. In Ref.\cite{rahmonov2016} phase dynamics of the SQUID with TNB was investigated. It was shown that this resonance branch is shifted by a factor of $\sqrt{2}$~\cite{rahmonov2016}. An interesting problem concerns the effect of external electromagnetic radiation. Its well known that under influence of the external radiation in the IV--curve appears Shapiro step (SS) and its subharmonics due to frequency locking by the external radiation. Their positions on IV-curve depend on the radiation frequency, and whose widths depend on the radiation frequency and amplitude. On the other hand, IV--curve of SQUID has the resonance branch due to beating states~\cite{schmidt85}. So it is interesting to investigate the effect of external radiation and external magnetic field on the dynamics of SQUID. The properties of SS on the resonance branch are not investigated yet, up to now. Particularly, the amplitude dependence of the SS's width on the resonance branch of the SQUID has not been calculated yet.

In this paper we investigate the effect of external radiation and dc magnetic field on the dynamics of the DC-SQUID with nontrivial barriers. We perform its comparative analysis with the trivial case. We analyze the behavior of Shapiro step and its second harmonic with the changing of the amplitude of the external radiation. In section I are considered the theoretical model and system of equations, which describes the dynamics of DC--SQUID. In section II is shown the results of numerical simulation and their discussions. In conclusion is summarized the obtained results.

\section{Theoretical model and equations}

As was mentioned above, the presence of Majorana fermions leads to the single electron tunneling, which doubles period of the order parameter~\cite{veldhorst12}.

Consequently, to describe this system in the framework of RCSJ--model, it is enough to change $2e$ to $e$ and $\varphi$ to the $\varphi/2$. So, in both cases the Josephson relation is the same, i.e.
\begin{equation}
	\label{Jos_rel}
	\frac{\hbar}{e}\frac{d( \varphi/2)}{dt}=\frac{\hbar}{2e}\frac{d \varphi}{dt}=V
\end{equation}

\noindent where $\varphi$ and $V$ are phase difference and voltage of the JJ. The sum of currents for each JJ in the SQUID can be written in following form:

\begin{equation}
	\label{system_eq1}
	\left\{\begin{array}{ll}
		\displaystyle I_{1}=\frac{C\hbar}{2e}\frac{\partial^{2}\varphi_{1}}{\partial t^{2}}+\frac{\hbar}{2eR}\frac{\partial \varphi_{1}}{\partial t}+ I_{c}\sin(\frac{\varphi_{1}}{2})
		\vspace{0.3 cm}\\
		\displaystyle I_{2}=\frac{C\hbar}{2e}\frac{\partial^{2} \varphi_{2}}{\partial t^{2}}+\frac{\hbar}{2eR}\frac{\partial \varphi_{2}}{\partial t}+I_{c}\sin(\frac{\varphi_{2}}{2})
	\end{array}\right.
\end{equation}
\noindent where $C$ is capacitance, $R$ is resistance and $I_{c}$ is critical current of JJ, $I_{1}$ and $I_{2}$ are the currents passing through the JJs of the SQUID. We note that here is considered the effect of external electromagnetic radiation on the phase dynamics of SQUID. This effect is produced by the term $A\sin\omega t$, where $\omega$ is the frequency and $A$ is amplitude of harmonic current created by the radiation. The sum of total current through the system can be written as $I_{1}+I_{2}=I+Asin\omega t$. Note that in the system (\ref{system_eq1}) we only take into account the tunneling {through Majorana states.}
Actually, of course, there are also the standard $2\pi$--periodic Josephson current. In particular, in Ref.~\cite{dominguez12} (see. supplement materials, equations (37) and (41)) IV-curve both terms are investigated.

In oder to distinguish the main effects, we investigate the case with tunneling through Majorana states.
In the presence of external magnetic field a magnetic flux through the circuit is quantized
\begin{equation}
	\displaystyle \frac{1}{2 \pi}\frac{\varphi_1-\varphi_2}{2}+\frac{\varPhi_t}{\varPhi_0}=n
	\label{flux_quantum}
\end{equation}
\noindent where $\varPhi_{0}=h/2e$ is the magnetic flux quantum. The total flux $\varPhi_{t}$ through the SQUID can be determined by the expression
\begin{equation}
	\label{total_flux}
	\displaystyle \varPhi_t=\varPhi_{ext}+L I_{c} \sin(\frac{\varphi_{1}}{2}) -L I_{c} \sin(\frac{\varphi_{2}}{2})
\end{equation}
\noindent where $\varPhi_{ext}$ is a flux created by the external magnetic field, $L$ is an inductance of wires.

Using the Josephson relation (\ref{Jos_rel}), the expressions for  the currents (\ref{system_eq1}), the flux quantization (\ref{flux_quantum}), and total flux through the SQUID (\ref{total_flux}), we can write the system of equations in normalized units that describes the dynamics of SQUID

\begin{widetext}
\begin{equation}
	\label{eq_2}
	\left\{\begin{array}{ll}
		\displaystyle \frac{\partial \varphi_{1}}{\partial t}=V_{1}
		\vspace{0.3 cm}\\
		\displaystyle \frac{\partial V_{1}}{\partial t}=\frac{1}{\beta_{c}}\bigg\{\frac{I+A\sin\omega t}{2}
		-V_{1}-\sin(\frac{\varphi_{1}}{2})
		+\frac{1}{2\beta_{L}}\bigg[2\pi( n-\varphi_{ext})
		-\frac{\varphi_{1} - \varphi_{2}}{2}\bigg]\bigg\}
		\vspace{0.3 cm}\\
		\displaystyle \frac{\partial \varphi_{2}}{\partial t}=V_{2}
		\vspace{0.3 cm}\\
		\displaystyle \frac{\partial V_{2}}{\partial t}=\frac{1}{\beta_{c}}\bigg\{
		\frac{I+A\sin\omega t}{2}-V_{2}-\sin(\frac{\varphi_{2}}{2})
		-\frac{1}{2\beta_{L}}\bigg[2\pi( n-\varphi_{ext})
		-\frac{\varphi_{1} -\varphi_{2}}{2}\bigg]\bigg\}
	\end{array}\right.
\end{equation}
\end{widetext}

\noindent where $\beta_c=2 \pi I_c R^2 C /\varPhi_0$ is the McCumber parameter, $\beta_L =2 \pi L I_c/ \varPhi_0$ is the dimensionless inductance, and $\varphi_{ext}=\varPhi_{ext}/\varPhi_{0}$ is the normalized flux created by the external magnetic field. In the system of equations (\ref{eq_2}) time is normalized to $\omega_{c}=2eI_{c}R/\hbar$, the voltage is normalized to $V_{c}=I_{c}R$. The bias current $I$ and the amplitude of external radiation are normalized to $I_{c}$.

In the SQUID, the capacitances of JJs and the inductances of the wires form the oscillatory circuit, in which electromagnetic oscillations appear with the frequency

\begin{equation}
	\label{res_frequency}
	\omega_{b}=1/\sqrt{\beta_c \beta_L}
\end{equation}

Under the condition of $\omega_{J}=m\omega_{res}$, where $m$ is integer, an additional branch  appears on the IV-curve. The origin of it is connected to the resonance of Josephson oscillations and electromagnetic oscillations~\cite{schmidt85}.
For simplification in the following we say trivial (or nontrivial) SQUID, if the DC-SQUID is formed by using JJs with trivial (or nontrivial) barriers.

\section{Results and discussions}

First, of all we discuss briefly the properties of the DC--SQUID without external radiation.
\begin{figure}[h!]
 \centering
 \includegraphics[height=70mm]{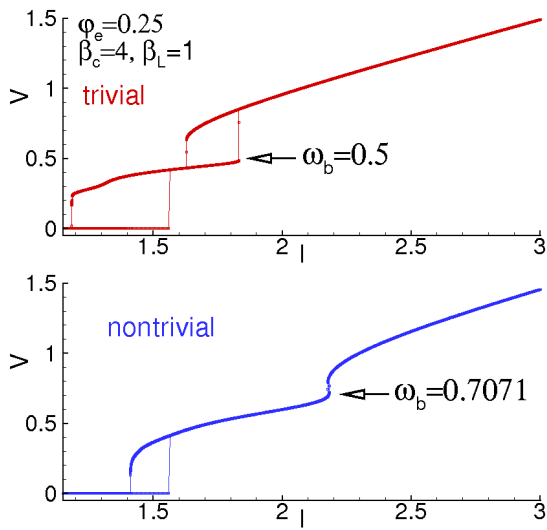}
 \caption{(a) IV-curve for trivial SQUID with parameters $\beta_{c}=4$ and $\beta_{L}=1$ in the external dc magnetic field $\varphi_{e}=0.25$; (b) The same as (a) for nontrivial SQUID.}
\label{iv_curves}
\end{figure}
As it was demonstrated in Ref.~\cite{schmidt85}, the presence of dc magnetic field leads to a branch (beating solution) in the IV--curve. The corresponding IV--characteristics for trivial SQUID is presented in Fig.\ref{iv_curves}(a). It is calculated by solving the system of equations (\ref{eq_2}) by the increasing and decreasing of bias current. The origin of this branch is a resonance between the Josephson and resonance circuit oscillations. It is satisfied by the frequency $\omega_{b}=1/\sqrt{\beta_{c}\beta_{L}}=0.5$. In Ref.~\cite{rahmonov2016} was shown that in case of nontrivial SQUID position of resonance branch of IV-curve shifts by $\sqrt{2}$. It can be seen in Fig.\ref{iv_curves}(b), which demonstrates the IV--curve of nontrivial SQUID calculated with the same parameters as Fig.\ref{iv_curves}(a), where the position of resonance branch corresponds to $\sqrt{2}\omega_{b}=\sqrt{2/\beta_{c}\beta_{L}}=0.7071$.

\begin{figure}[h!]
 \centering
 \includegraphics[height=70mm]{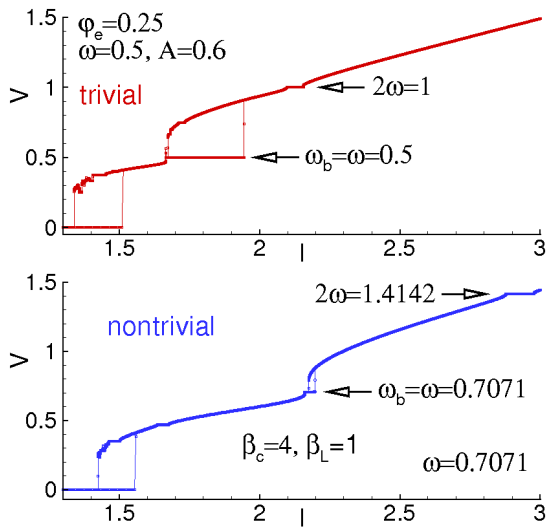}
 \caption{(a) IV--curve for trivial SQUID with parameters $\beta_{c}=4$ and $\beta_{L}=1$ at the external dc magnetic field $\varphi_{e}=0.25$ under external electromagnetic radiation $\omega=0.5$ and amplitude $A=0.6$; (b) The same as (a) for nontrivial SQUID at $\omega=0.7071$.}
\label{iv_curves_rad}
\end{figure}

Now we discuss the effect of external electromagnetic radiation. Fig.~\ref{iv_curves_rad}(a) represents the IV--characteristic of the trivial SQUID at $A = 0.6$ and $\omega=0.5$, which corresponds to the $\omega_{b}=\sqrt{1/\beta_{c}\beta_{L}}$. One can see, that IV--curve under radiation demonstrates a Shapiro step at $V = \omega = 0.5$ and some harmonics and subharmonics. We note that, here we investigate a behavior of main SS and its second harmonic, positions, which shown by the vertical arrows in Fig.~\ref{iv_curves_rad}(a). It can be seen that as usual the main SS width is visually larger than the width of its second harmonic. Figure~\ref{iv_curves_rad}(b) demonstrates IV--curve in case of nontrivial SQUID at $A=0.6$ and $\omega=0.7071$. This figure demonstrates that the width of second harmonic is visually larger than main SS width.

\begin{figure}[h!]
 \centering
 \includegraphics[height=60mm]{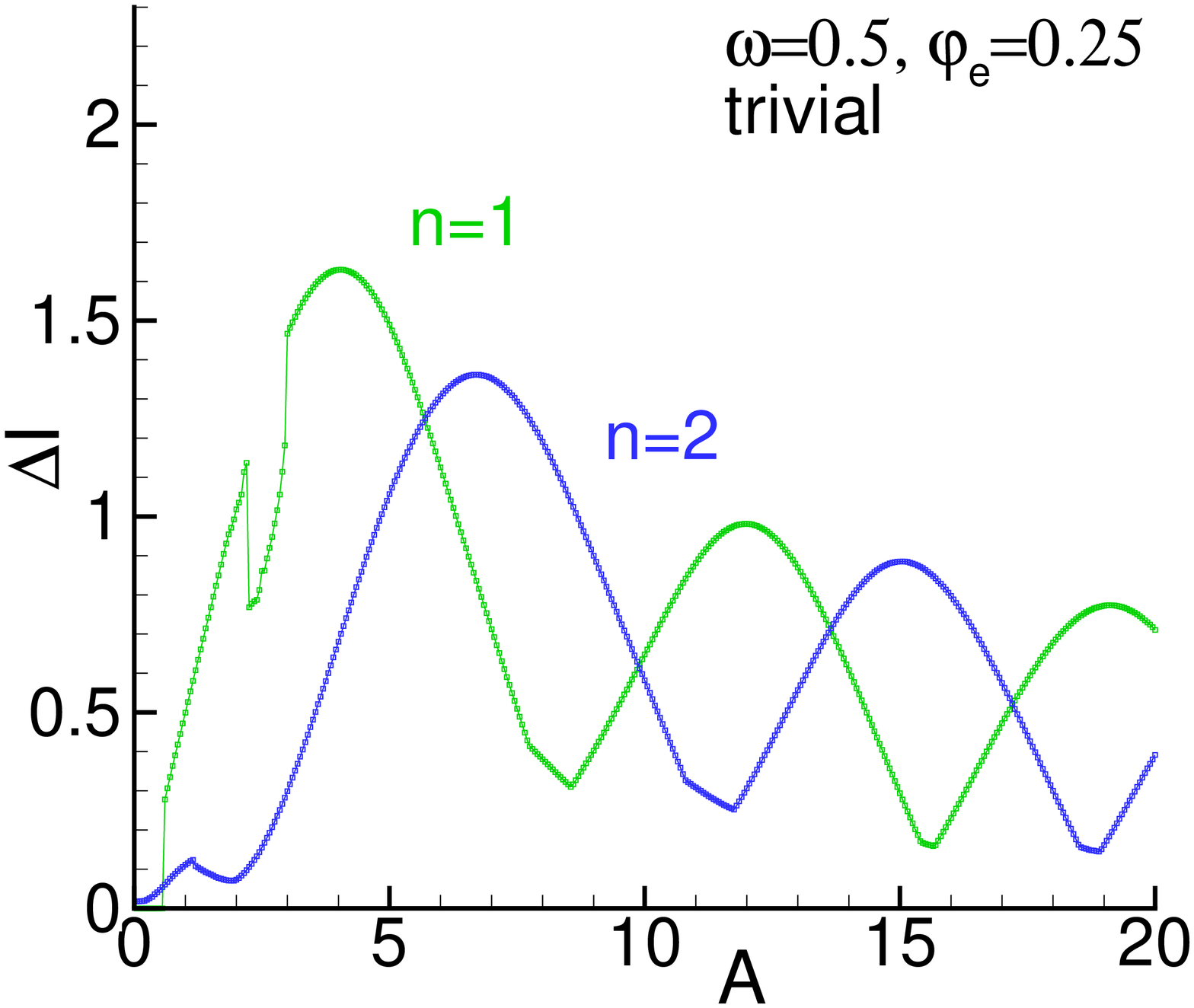}
  \includegraphics[height=60mm]{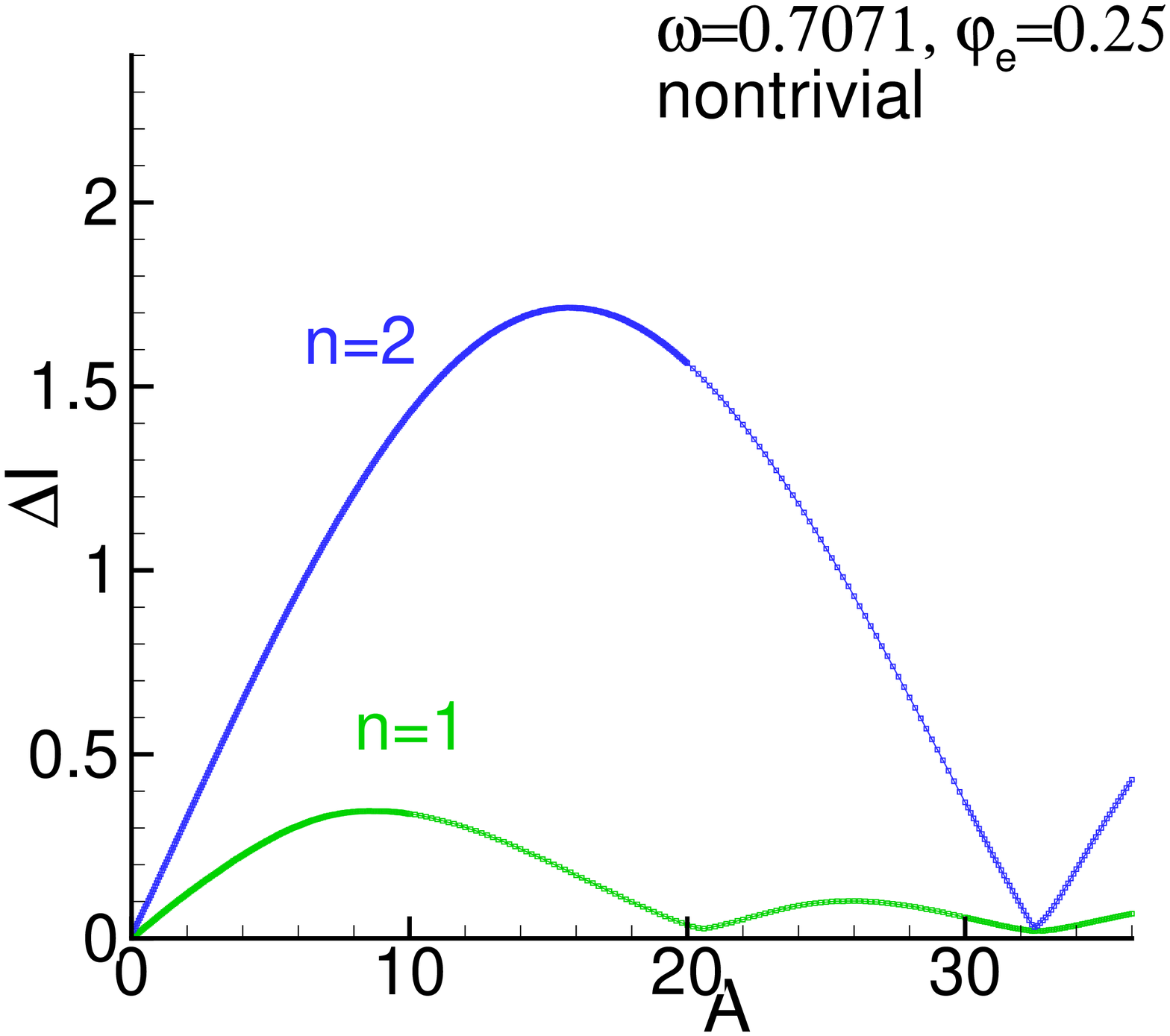}
 \caption{(a) The amplitude dependence of main Shapiro step and its second harmonic for $\omega=0.5$ at $\varphi_{e}=0.25$ for the SQUID with trivial barriers;
 (b) The same as the case (a) for SQUID with nontrivial barriers calculated for $\omega=0.7071$ at $\varphi_{e}=0.25$.}
\label{ss_amp_dep_phe0-25}
\end{figure}

To investigate SS and its harmonics width behavior in detail, we have calculated its amplitude dependence for each considered cases. Here let us discuss the trivial case. Fig.\ref{ss_amp_dep_phe0-25}(a) shows the amplitude dependence of the width of both the main SS and its second harmonic for $\omega=0.5$ at the $\varphi_{e}=0.25$. In this case the main SS is located on the resonance branch of IV--curve, which corresponds to the $\omega_{b}=0.5$. This figure demonstrates that the first maximum of the main SS is located at $A=4.05$ and its width is $\Delta I=1.63$. This is wider than the first maximum of the second harmonic which is equal to $\Delta I=1.362$ and corresponds to $A=6.7$. We note that in the amplitude interval corresponding to $[2.2, 3]$ the width of the main SS is reduced due to chaotic behavior~\cite{kautz85}.

In nontrivial case the position of resonance branch is shifted by factor of $\sqrt{2}$~\cite{}. So, to describe the effect of resonance on SS we should choose the frequency of radiation as $\omega=0.5\sqrt{2}=0.7071$. Fig.~\ref{ss_amp_dep_phe0-25}(b) shows the amplitude dependence of main SS and its second harmonic in the presence of resonance. Here the maximal of width of second harmonic is equal to $\Delta I=1.714$ and corresponds to the amplitude $A=15.8$. It is larger than main SS width, which is equal to $\Delta I=0.3465$ (at $A=8.55$). So the comparative analysis of the main SS and its second harmonics allows to detect the Majorana fermions.

\begin{figure}[h!]
 \centering
 \includegraphics[height=60mm]{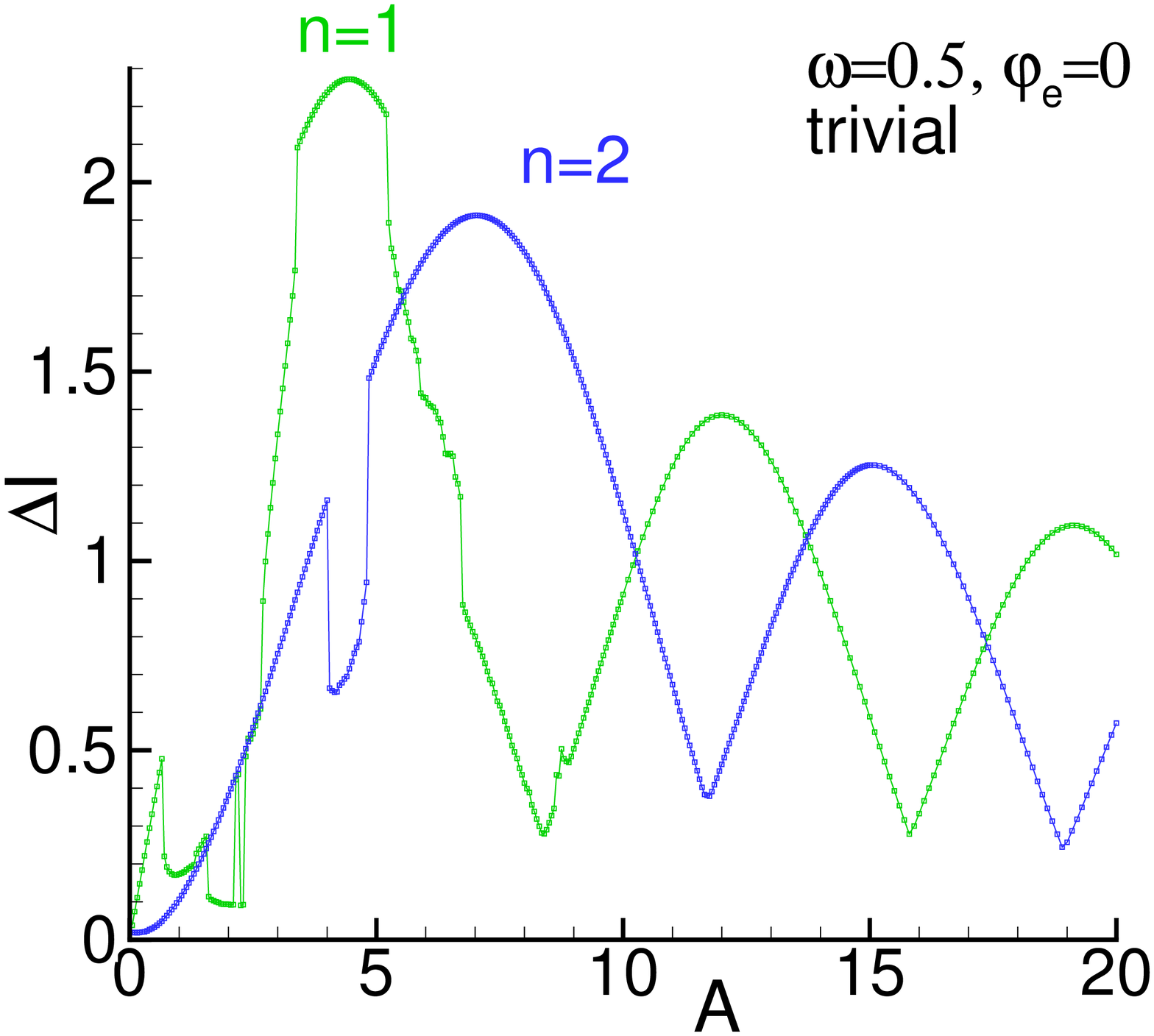}
 \includegraphics[height=60mm]{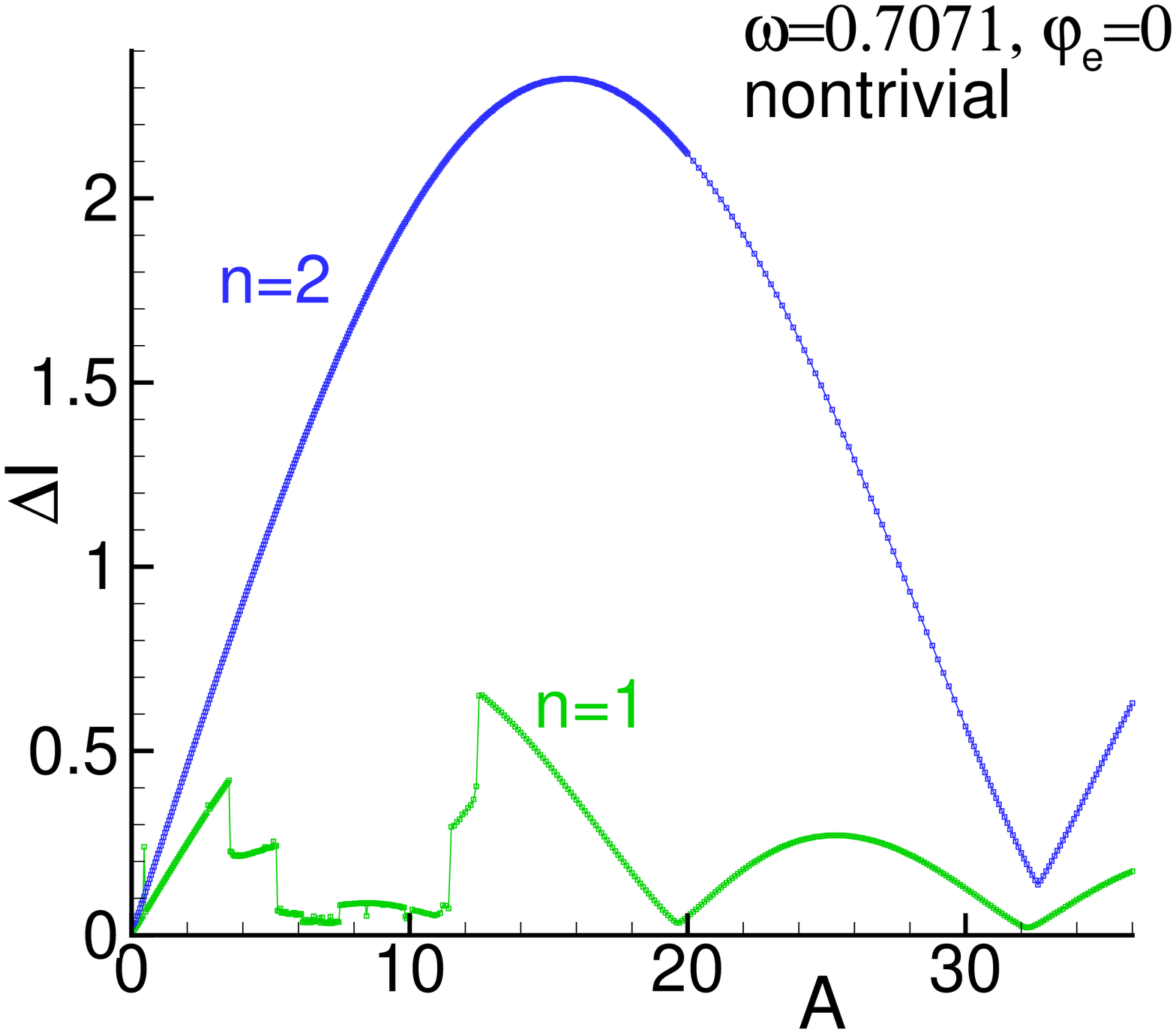}
 \caption{(a) The amplitude dependence of main Shapiro step and its second harmonic for $\omega=0.5$ in the absence of magnetic field for SQUID with trivial barriers; (b) The same as (a) for $\omega=0.7071$ for SQUID with nontrivial barriers.}
\label{ss_amp_dep_phe0}
\end{figure}

To distinguish the effect of resonance let us compare the above discussed case with a case of the absence of resonance, which corresponds to zero external magnetic field. The amplitude dependence of widths of main SS and its second harmonic for trivial SQUID is demonstrated in Fig.~\ref{ss_amp_dep_phe0}(a). One can see that the first maxima of the main SS (which is $\Delta I =2.272$ corresponding to $A=4.45$) and its second harmonic (which is $\Delta I =1.9125$ corresponding to $A=7$) is larger than the corresponding points with a resonance. Based on this fact we can conclude that the presence of resonance leads to the reduction of period of amplitude dependence of the width. An another interesting difference is in case of absence of resonance there are many regions (in intervals) with the reduced main SS width, which corresponds to the chaotic behavior of IV-curve. So the presence of resonance will reduce the chaotic behavior of IV--curve. The amplitude dependence of SS width and its second harmonic in the absence of resonance for nontrivial SQUID is shown Fig.~\ref{ss_amp_dep_phe0}(b). The main difference in this case is that here the qualitatively the behavior of second harmonic is like main SS behavior in trivial SQUID.

\section{Conclusion}
The influence of the external electromagnetic radiation on the IV--curves of DC--SQUID with topologically trivial and nontrivial barriers in the presence of dc magnetic field were studied. The amplitude dependence of Shapiro step width is investigated. It is found that for a trivial DC--SQUID in the absence of resonance branch the maximal width of Shapiro step and periods of its amplitude dependence increased  in compare with a resonance case.
For the nontrivial DC--SQUID the second harmonic is wider than that of the corresponding trivial case, and the main Shapiro step has a smaller width than that of the corresponding trivial one.
We consider that the measurement of the Shaprio step and its second harmonic with changing the amplitude of the external electromagnetic radiation can be used  as a tool for detection of the  Majorana fermions.

\section{Acknowledgments}
We thank ... for detailed discussion of this paper. The reported study was funded by RFBR according to the research projects 15--51--61011\_Egypt, 15--29--01217 and 16--52--45011\_India.

{}

\end{document}